\documentclass[twocolumn]{aastex63}

\usepackage[utf8]{inputenc} 
\usepackage{amsmath, amsbsy}
\usepackage{hyperref} 

\shorttitle{Korg: 1D LTE spectral synthesis}
\shortauthors{Wheeler et al.}

\newcommand{\kcc}{K_{\mathrm{C}_2}}
\newcommand{\dd}[1]{\textrm{d}#1}
\newcommand{\codename}[1]{\textsc{#1}}
\newcommand{\korg}{\codename{Korg}}
\newcommand{\moog}{\codename{Moog}}
\newcommand{\ts}{\codename{Turbospectrum}}
\newcommand{\sme}{\codename{SME}}

\newcommand{\newchange}[1]{#1}
\newcommand{\change}[1]{#1}

\begin{document}
\title{Korg: a modern 1D LTE spectral synthesis package}

\author[0000-0001-7339-5136]{Adam J. Wheeler}
\affiliation{Department of Astronomy, Ohio State University, McPherson Laboratory, 140 West 18th Avenue, Columbus, Ohio}
\author[0000-0002-7918-3086]{Matthew W. Abruzzo}
\affiliation{Department of Astronomy, Columbia University, Pupin Physics Laboratories, New York, NY 10027, USA}
\author[0000-0003-0174-0564]{Andrew R. Casey}
\affiliation{School of Physics \& Astronomy, Monash University, Victoria, Australia}
\affiliation{Center of Excellence for Astrophysics in Three Dimensions (ASTRO-3D)}
\author[0000-0001-5082-6693]{Melissa K. Ness}
\affiliation{Department of Astronomy, Columbia University, Pupin Physics Laboratories, New York, NY 10027, USA}

\correspondingauthor{Adam Wheeler}
\email{wheeler.883@osu.edu}

\begin{abstract}
\noindent
We present \korg{}, a new package for 1D LTE (local thermodynamic equilibrium) spectral synthesis of FGK stars, which computes theoretical spectra from the near-ultraviolet to the near-infrared, and implements both plane-parallel and spherical radiative transfer.
We outline the inputs and internals of \korg{}, and  compare synthetic spectra from \korg{}, \moog{}, \ts{}, and \sme{}.
The disagreements between \korg{} and the other codes are no larger than those between the other codes, although disagreement between codes is substantial.
We examine the case of a C$_2$ band in detail, finding that uncertainties on physical inputs to spectral synthesis account for a significant fraction of the disagreement.
\korg{} is 1--100 times faster than other codes in typical use, compatible with automatic differentiation libraries, and easily extensible, making it ideal for statistical inference and parameter estimation applied to large data sets.
Documentation and installation instructions are available at \url{https://ajwheeler.github.io/Korg.jl/stable/}.  
\end{abstract}

\keywords{spectroscopy}

\section{Introduction}
Improvements in instrumentation have yielded exponential growth in the amount of spectral data to analyse.
Creating pipelines that can keep up with analysis is nontrivial.
There are several extant codes for 1D LTE spectral synthesis, including \ts{} \citep{plezLithiumAbundancesOther1993, plezTurbospectrumCodeSpectral2012, gerberNonLTERadiativeTransfer2022a}, \moog{} \citep{snedenNitrogenAbundanceVery1973, snedenMOOGLTELine2012}, \codename{SYNTHE} \citep{kuruczSYNTHESpectrumSynthesis1993, sbordoneATLASSYNTHELinux2004}, \codename{SME}, \citep{valentiSpectroscopyMadeEasy1996, valentiSMESpectroscopyMade2012, piskunovSpectroscopyMadeEasy2017, wehrhahnPySMESpectroscopyMade2022}, \codename{SPECTRUM} \citep{grayCalibrationMKSpectral1994}, and \codename{SYNSPEC} \citep{hubenySynspecGeneralSpectrum2011, hubenyBriefIntroductoryGuide2017, hubenyTLUSTYSYNSPECUsers2021a}.
While they have enabled a huge volume of research, these codes can be difficult to use for the uninitiated, and require input and output through custom file formats, impeding integration into analysis code.
Here we present \korg, a new 1D LTE synthesis package, written in \codename{Julia} and suitable for easy integration with scripts and use in an interactive environment.
\change{As the first such new code in more than two decades, \korg{} benefits from numerical libraries not available at the time earlier packages were authored, principally modern automatic differentiation packages and optimization libraries.}

\change{
The two fundamental assumptions made by \korg{} are that the stellar atmosphere is hydrostatic and 1D, and in LTE (local thermodynamic equilibrium).
Eliminating the former two assumptions means calculating atmospheric structure using the full hydrodynamic equations (e.g. \citealp{freytagSimulationsStellarConvection2012, magicStaggergridGrid3D2015, schultzSynthesizingSpectra3D2022}), or the magnetohydrodynamic equations (e.g. \citealp{voglerSimulationsMagnetoconvectionSolar2005}), if the internal magnetic field is strong.
}
Fortunately, corrections to 1D LTE level populations (e.g. \citealp{amarsiGALAHSurveyNonLTE2020, amarsi3DNonLTEIron2022}), equivalent widths, or abundances (e.g. \citealp{lindNonLTECalculationsNeutral2011, amarsiGalacticChemicalEvolution2015, bergemannNLTEAnalysisSr2012, osorioMgLineFormation2016}) can be calculated from NLTE simulations.
These can be applied to LTE results or codes to produce approximate NLTE spectra at relatively little computational cost.
Additionally, biases from the assumption of LTE  will roughly cancel for similar stars, yielding differential abundance estimates with high precision.

\begin{figure}
    \centering
    \includegraphics{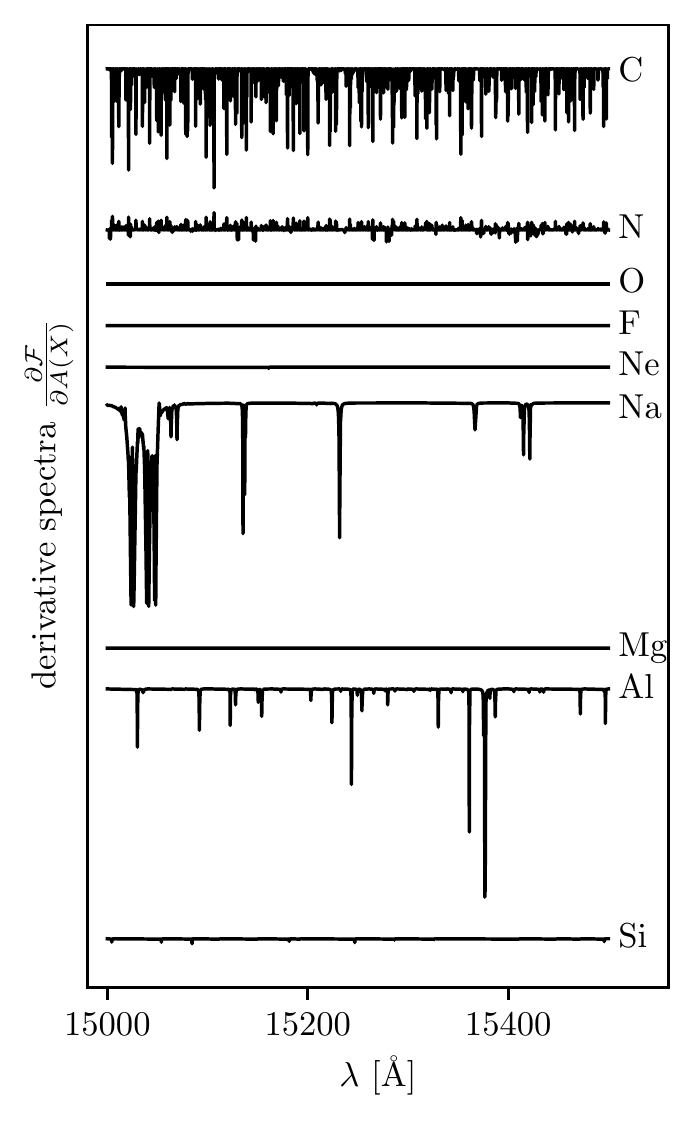}
    \caption{Derivatives of a synthesized solar spectrum with respect to various abundances, $\frac{\partial \mathcal{F}}{\partial A(X)}$. Vertical offsets have been applied, but the relative scaling is preserved, which is why some derivative spectra appear flat here (but show fluctuations at a sub-percent level).}
    \label{fig:gradients}
\end{figure}

The performance of spectral synthesis codes is most important when fitting observational data.
Because synthesis must be embedded in an inference loop, the analysis of of a single spectrum may trigger tens or hundreds of syntheses.
Even when many parameters may be estimated by interpolating over (or otherwise comparing to) a precomputed grid of spectra (e.g. \citealp{recio-blancoAutomatedDerivationStellar2006,smiljanicGaiaESOSurveyAnalysis2014, holtzmanAPOGEEDataReleases2018, boecheSPAceV12021, buderGALAHSurveyThird2021}), individual abundances are best done with targeted syntheses of individual lines. Furthermore, generating a grid from which to interpolate can be computationally expensive.
Inference and optimization can also be sped up by fast and accurate derivatives of the function being sampled or minimized, most easily produced via automatic differentiation. 
\korg{} is designed to be compatible with automatic differentiation packages (e.g. \textsc{ForwardDiff}), which can provide derivative spectra in roughly the same amount of time required for a single synthesis (as discussed in Section \ref{sec:benchmarks}; see Figure \ref{fig:gradients}).

\section{Description of code}
To synthesize a spectrum, \korg{} calculates the number density of each species (e.g. H I, C II, CO) at each layer in the atmosphere (Section \ref{sec:moleq}), then computes the absorption coefficient at each wavelength and atmospheric layer due to continuum (Section \ref{sec:continuum}) and line absorption (Section \ref{sec:lines}).
Given the total absorption coefficient at each wavelength and atmospheric layer, it then solves the radiative transfer equation to produce the flux at the stellar surface (Section \ref{sec:radiative_transfer}).

\subsection{Inputs}
\korg{} takes as inputs, a model atmosphere, a linelist, and abundances for each element in $A(X)$ form,\footnote{$A(X) = \log_{10}(n_X/n_\text{H}) + 12$ where $n_X$ is the total number density of element $X$ and $n_\text{H}$ that of hydrogen.} assumed to be constant throughout the atmosphere.
\korg{} includes functions for parsing \texttt{.mod}-format MARCS \citep{gustafssonGridMARCSModel2008} model atmospheres, and line lists in the format supplied by VALD \citep{piskunovVALDViennaAtomic1995, kupkaVALD2ProgressVienna1999, kupkaVALD2NewVienna2000, 2015PhyS...90e4005R, pakhomovHyperfineSplittingVALD2019} or Kurucz \citep{2011CaJPh..89..417K}, or those accepted by \moog.
When parsing VALD linelists, \korg{} will automatically apply corrections to unscaled $\log(gf)$ values for transitions with isotope information using the isotopic abundance values supplied by \citet{berglundIsotopicCompositionsElements2011} for linelists which are not already adjusted for isotopic abundances.
\korg{} detects and uses packed Anstee, Barklem, and O'Mara (ABO; \citealp{ansteeInvestigationBruecknerTheory1991, barklemListDataBroadening2000}) if they are provided (as VALD optionally does).
It uses vacuum wavelengths internally, and will automatically convert air-wavelength VALD linelists to vacuum \citep{birchCorrectionUpdatedEdlen1994}.
Users may also construct line lists or model atmospheres using custom code and pass them directly into \korg{}.

\subsection{Chemical equilibrium} \label{sec:moleq}

\korg{} uses the assumption of LTE, i.e. that in a sufficiently small region, baryonic matter is described by thermal distributions, and radiation is only slightly out of detailed balance.
The source function (the ratio of the per-volume emission and absorption coefficients) is dominated by collisions of baryonic matter and is Planckian.
While non-LTE (NLTE) calculations are important for producing the most unbiased possible model spectra, they are prohibitively slow, and not yet suitable for applications that require computing many spectra over large wavelength ranges.

In order to compute the absorption coefficient, $\alpha_\lambda$, at each layer of the atmosphere, \korg{} must solve for the number density of each species.
Treating the number density of each neutral atomic species as the free parameters, \korg{} uses the \codename{NLsolve} library \citep{mogensenJuliaNLSolversNLsolveJl2020} to solve the system of Saha and molecular equilibrium equations with the temperature, total number density, and number density of free electrons set by the model atmosphere, and the total abundance of each element set by the user.
By default, \korg{} uses molecular equilibrium constants from \citet{barklemPartitionFunctionsEquilibrium2016} (the ionization energies are originally from \citealp{haynesCRCHandbookChemistry2010}), but alternatives can be passed by the user.
\change{
These are defined only up to 10,000\,K, so we treat the molecular partition functions as constant above this temperature. 
This is unlikely to be problematic as few molecules are present above this threshold.
The default atomic partition functions are calculated using energy levels from NIST\footnote{\url{https://physics.nist.gov/PhysRefData/ASD/levels_form.html}} \citep{kramidaNISTAtomicSpectra2021}.
}

\begin{figure}
    \centering
    \includegraphics[width=0.45\textwidth]{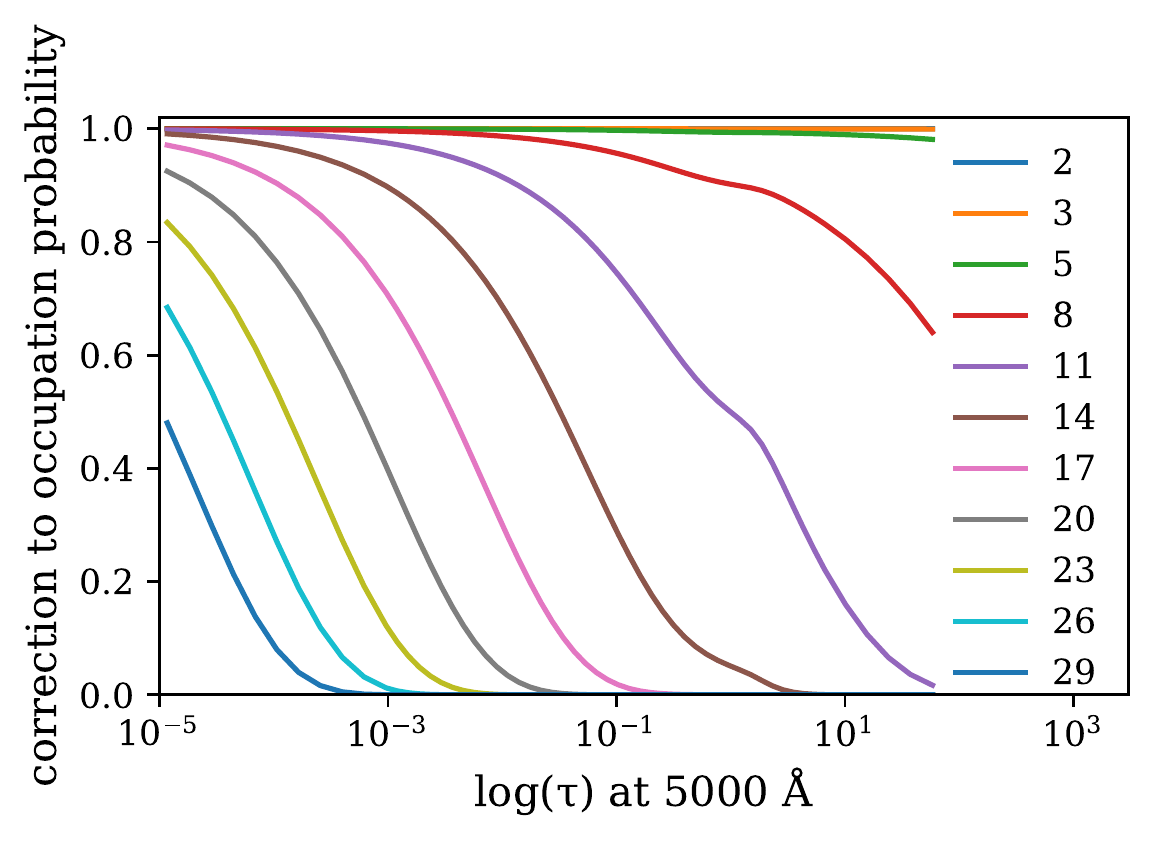}
    \caption{The \citet{hummerEquationStateStellar1988} occupation probability correction factor, $w$, for several values of the primary quantum number, $n$, in hydrogen, evaluated  at each layer in the solar atmosphere (indexed by optical depth at 5000\,\AA). While the formalism affects transitions involving outer energy levels, those which don't, e.g. $H_\alpha$, are unaffected. The results are similar for the other atmospheres considered in Section \ref{sec:verification}.
    }
    \label{fig:winatm}
\end{figure}

\change{
In dense environments, upper energy levels are perturbed or dissolved.
This can be crudely accounted for by truncating the terms of the partition function (e.g. \citealp{hubenyTheoryStellarAtmospheres2014} section 4.1), or with the ``probability occupation formalism'' developed in \citet{hummerEquationStateStellar1988}, and  generalized by \citet{hubenyNLTEModelStellar1994}.
This effect is most important for hydrogen, where it strongly impacts the partition function starting above 10,000\,K.
Based on energy level truncation, the partition functions of lithium and vanadium are 
affected in the same regime at the 5\% and 3\% level, respectively, but 
other elements are unchanged at the 1\% level.
For FGK stars, these effects are most important for higher-order hydrogen transitions.
Figure \ref{fig:winatm} shows the occupation probability correction factor, $w$ for several energy levels of hydrogen in the solar atmosphere.
For $n=1,2,3$, the correction is smaller than $10^{-3}$ everywhere.
At present, we do not include these effects in \korg{}, because they do not make a large difference for FGK stars, and because of disagreement with observational data (see Section \ref{sec:balmer}).
}

By default, \korg{} solves the equilibrium equations taking into account all elements up to uranium \newchange{(as neutral, singly ionized, and doubly ionized species)}, as well as 247 diatomic molecules.
As they exist in extremely small quantities in LTE atmospheres, \korg{} neglects species which are more than doubly ionized.
It also neglects ionic and triatomic molecules, \change{which are present in cool stars}, although we plan to address this in a future version.
\subsection{Continuum absorption} \label{sec:continuum}
\begin{table*}[]
    \centering
    \begin{tabular}{lllp{6cm}}
        \textbf{Absorption source} & \textbf{Wavelength bounds} & \textbf{Temperature bounds} & \textbf{Reference} \\
        \hline
        H$^-$ bf & $\geq 1250$ \AA{} & & \citet{mclaughlinLessSupGreater2017} \\
        H$^-$ ff & $2604$ \AA{} -- $113,918$ \AA& 2520 K -- 10,080 K &  \citet{bellFreefreeAbsorptionCoefficient1987}\\
        He$^-$ ff & 5063 \AA{} -- 151,878 \AA{} & 2520 K -- 10,080 K & \citet{1994MNRAS.269..871J} \\
        H I bf & & &  \citet{1961ApJS....6..167K} via \citet{gingerichTheoryObservationNormal1969} via \citet{1970SAOSR.309.....K} \\
        H I ff & 100 \AA{} -- $10^6$ \AA{} & $100$ K -- $10^6$ K &  \citet{vanhoofAccurateDeterminationFreefree2014} \\
        He II bf &&& \citet{1961ApJS....6..167K} via \citet{gingerichTheoryObservationNormal1969} via \citet{1970SAOSR.309.....K} \\
        H$_2^+$ bf and ff & 700 \AA{} -- 20,000 \AA{} & 3150 K -- 25,200 K & \citet{stancilContinuousAbsorptionHe1994} \\
        He II ff &100 \AA{} -- $10^6$ \AA{} & $100$ K -- $10^6$ K &  \citet{vanhoofAccurateDeterminationFreefree2014}  with departure coefficients from \citet{peachContinuousAbsorptionCoefficients1970}\\
        metals ff & 100 \AA{} -- $10^6$ \AA{} & 100 K -- $10^6$ K & \citet{vanhoofAccurateDeterminationFreefree2014} with departure coefficients from \citet{peachContinuousAbsorptionCoefficients1970} for C I, Si I, and Mg I \\
        C I bf & 500 \AA{} -- 30,000 \AA{} & 100 K   -- $10^5$ K & \citet{naharPhotoionizationElectronionRecombination1991}\\
        Na I bf & 500 \AA{} -- 30,000 \AA{} & 100 K  -- $10^5$ K & \citet{seatonOpacitiesStellarEnvelopes1994}\\
        Mg I bf & 500 \AA{} -- 30,000 \AA{} & 100 K  -- $10^5$ K & \citet{seatonOpacitiesStellarEnvelopes1994}\\
        Al I bf & 500 \AA{} -- 30,000 \AA{} & 100 K  -- $10^5$ K & \citet{seatonOpacitiesStellarEnvelopes1994}\\
        Si I bf & 500 \AA{} -- 30,000 \AA{} & 100 K  -- $10^5$ K & \citet{naharAtomicDataOpacity1993}\\
        S I bf & 500 \AA{} -- 30,000 \AA{} & 100 K   -- $10^5$ K & \citet{seatonOpacitiesStellarEnvelopes1994}\\
        Ca I bf & 500 \AA{} -- 30,000 \AA{} & 100 K  -- $10^5$ K & \citet{seatonOpacitiesStellarEnvelopes1994}\\
        Fe I bf & 500 \AA{} -- 26,088 \AA{} & 100 K  -- $10^5$ K & \citet{bautistaAtomicDataIRON1997} \\
        H Rayleigh & $\geq 1300$ \AA{} && \citet{leeExactLowenergyExpansion2005} via \citet{colganNewGenerationAlamos2016}\\
        He Rayleigh & $\geq 1300$ \AA{} && \citet{dalgarnoRefractiveIndicesVerdet1960, dalgarnoSpectralReflectivityEarth1962,schwerdtfegerAtomicStaticDipole2006} via \citet{colganNewGenerationAlamos2016}\\
        H$_2$ Rayleigh & $\geq 1300$ \AA{} && \citet{dalgarnoRayleighScatteringMolecular1962} \\
        electron scattering &&& \citet{thomsonXLIIIonizationMoving1912} \\
        \hline
    \end{tabular}
    \caption{Sources of continuum absorption in \korg{}.  When bounds in temperature or wavelength are missing, the absorption coefficient is defined for all positive values.}
    \label{tab:continuum_sources}
\end{table*}
\begin{figure}
    \centering
    \includegraphics[width=0.45\textwidth]{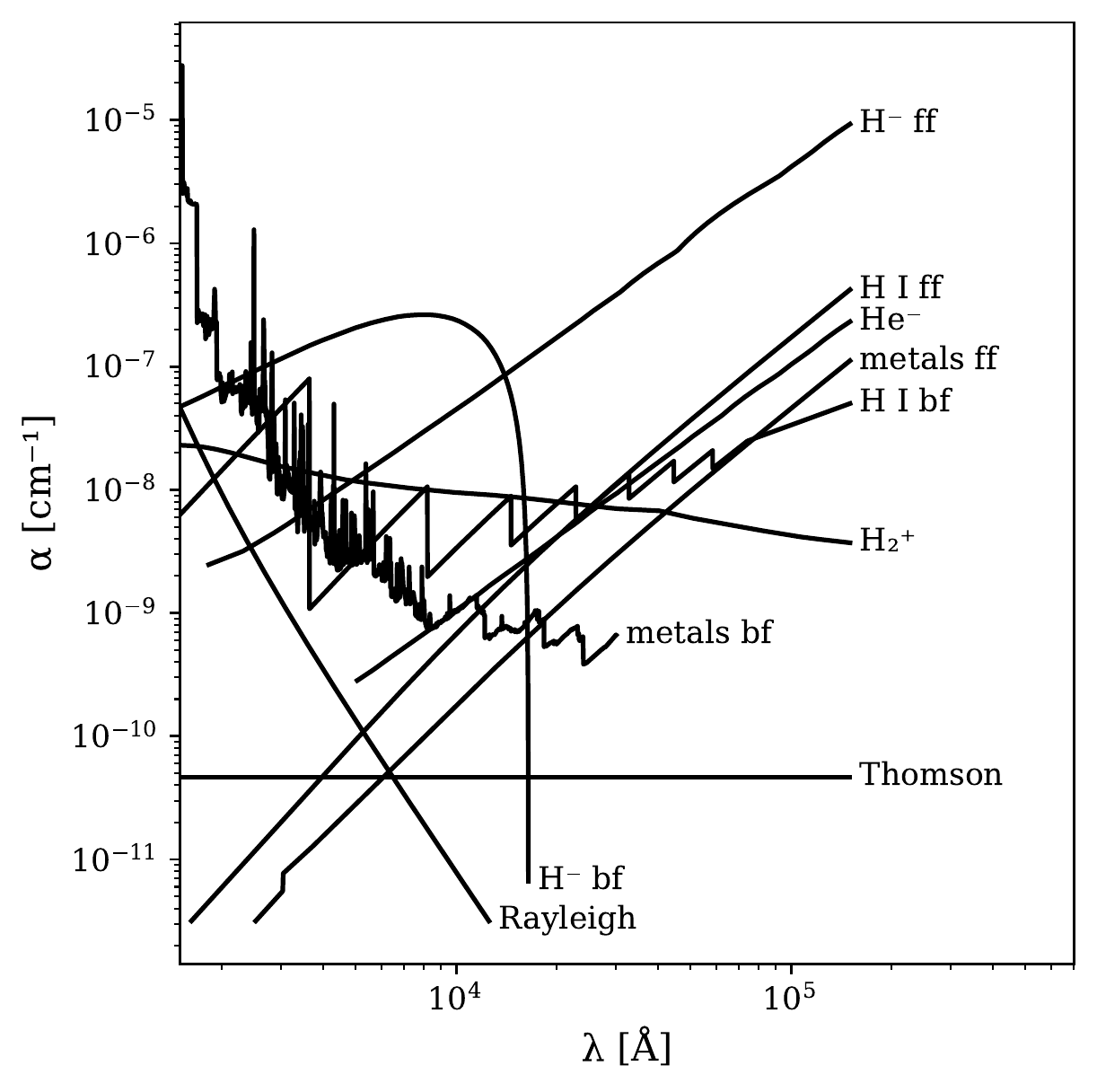}
    \caption{Major sources of opacity at the solar photosphere, defined here at the MARCs model atmosphere layers where the optical depth at 5000 \AA{} is most nearly unity.}
    \label{fig:abs}
\end{figure}
\korg{} computes contributions to the continuum absorption coefficient from a number of sources, listed in Table \ref{tab:continuum_sources}.
Continuum absorption mechanisms involving an atom and an electron can be classified as bound-free (bf) or free-free (ff), depending on whether the electron is initially bound to the atom.\footnote{Note that these are named as though the electron were bound, even for free-free interactions. 
For example, ``H I bf'' refers to the ionization of neutral hydrogen, and ``H I ff'' refers to absorption by the interaction of free electrons and free protons.}
\change{
In addition to bound-free and free-free absorption for H$^-$, H I, He II, and H$_2^+$ (as well as He$^-$ and He), \korg{} includes treatments of bound-free and free-free interactions with metals.
Free-free interactions are treated with the hydrogenic approximation using 
Gaunt factors from \citet{vanhoofAccurateDeterminationFreefree2014}, with corrections from \citet{peachContinuousAbsorptionCoefficients1970} for neutral He, C, Si, and Mg.
Bound-free interactions are treated with cross-sections from NORAD\footnote{\url{https://norad.astronomy.osu.edu/\#AtomicDataTbl1}}, when available, TOPBase\footnote{\url{https://cds.u-strasbg.fr/topbase/topbase.html}} \citet{seatonOpacitiesStellarEnvelopes1994} otherwise.
Following \citet{gustafssonGridMARCSModel2008}, we shifted the theoretical energy levels to the empirical values from NIST, leaving out levels for which an empirical counterpart wasn't present.
We considered all species with bound-free interaction included in \citet{gustafssonGridMARCSModel2008}, but included only those which contribute to the total continuum absorption at above the $10^{-3}$ level anywhere in any of the atmospheres used in Section \ref{sec:verification}.
Figure \ref{fig:abs} shows the contribution of various mechanisms to the continuum absorption coefficient at the solar photosphere.
}
At present, \korg{} treats all scattering as absorption, which is correct in the LTE regime.
In the future we plan to support quasi-LTE radiative transfer with isotropic scattering, which will yield more accurate spectra when Rayleigh scattering dominates or is a significant source of opacity.

\begin{figure}
    \centering
    \includegraphics[width=0.45\textwidth]{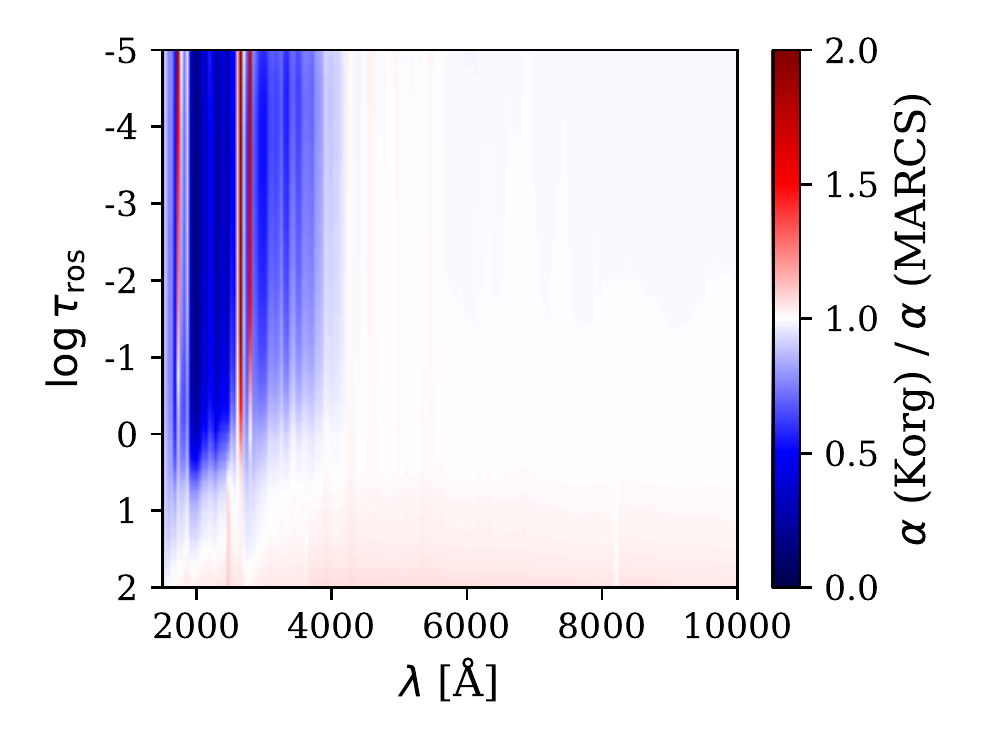}
    \caption{The ratio of the absorption coefficient per \korg{} and per MARCS in the solar atmosphere (indexed by Rosseland mean opacity, $\tau_\text{ros}$).  Agreement is good, except in the blue and ultraviolet where \korg{} uses more recent metal opacities and the effects of line blanketing (handled separately in \korg{}) in MARCS become large.}
    \label{fig:alpharatio}
\end{figure}

\change{
Figure \ref{fig:alpharatio} shows the ratio of the continuum absorption according to \korg{} and the one used by MARCS through the solar atmosphere.
Agreement is good, except in the violet and ultraviolet, where \korg{} uses more recent bound-free metal absorption coefficients with more structure in wavelength, which shows up as vertical bands in the figure.
Compounding this, the MARCS values also include the effects on line blanketing, which is handled via the linelist in \korg{}.
The agreement between the continua calculated by \korg{} and other codes is good (see Section \ref{sec:verification}).
}

\subsection{Line absorption} \label{sec:lines}
The contribution of each line to the total absorption coefficient, $\alpha_\text{line}$, is
\begin{equation}
    \alpha_\lambda^\text{line} = \sigma n \frac{\exp(-\beta E_\text{up}) - \exp(-\beta E_\text{lo})}{U(T)} \phi(\lambda)
\end{equation}
where the wavelength-integrated cross-section, $\sigma$ is given by
\begin{equation}
    \sigma = g_\text{lo} f \frac{\pi e^2 \lambda_0^2}{m_e c^2} \quad .
\end{equation}
Here, $n$ is the number density of the line species, $E_\text{up}$ and $E_\text{lo}$ are the upper and lower energy levels of the transition, $\beta = 1/kT$ is the thermodynamic beta, $\phi$ is the normalized line profile, $U$ is the species partition function, $T$ is the temperature, $f$ is the oscillator strength, $g_\text{lo}$ is the degeneracy of the lower level, $\lambda_0$ is the line center, and $m_e$ is the electron mass.

For all lines of all species besides hydrogen \change{(including autoionizing lines)}, $\phi$ is approximated with a Voigt profile using the numerical approximation from \citet{hungerZurTheorieWachstumskurven1956}.
The width of the Gaussian component is 
\begin{equation}
    \sigma_D = \lambda_0 \sqrt{\frac{2kT}{m} + \xi^2} \quad ,
\end{equation}
where $m$ is the species mass, and $\xi$ is the microturbulent velocity, a fudge factor used to account for convective Doppler-shifts.
The width of the Lorentz component (in frequency rather than wavelength units) is given by
\begin{equation}
    \Gamma = \Gamma_\text{rad} + \gamma_\text{stark} n_e + \gamma_\text{vdW}n_\text{H I} \quad ,
\end{equation}
where $\Gamma_\text{rad}$ is the radiative broadening parameter, $\gamma_\text{stark}$ the per-electron Stark broadening parameter, $\gamma_\text{vdW}$ the per-neutral-hydrogen van der Waals broadening parameter, and $n_e$ and $n_\text{H I}$ are the number densities of free electrons and neutral hydrogen, respectively.
We neglect pressure broadening of molecular lines, setting $\gamma_\text{stark}$ and $\gamma_\text{vdW}$ to zero.

When $\Gamma_\text{rad}$ is not supplied in the linelist, \korg{} approximates it with
\begin{equation}
    \Gamma_\textrm{rad} = \frac{8 \pi^2 e^2}{m c \lambda^2} gf \quad ,
\end{equation}
where $m$ is the mass of the atom or molecule, $g$ is the degeneracy of the lower level of the transition, and $f$ is the transition oscillator strength. 
This can be obtained by assuming that spontaneous de-exitation dominates the transition's energy uncertainty, and that the upper level's degeneracy is unity.

The values of $\gamma_\text{stark}$ and $\gamma_\text{vdW}$ at $10^5$ K, $\gamma^\text{stark}_0$ and $\gamma^\text{vdW}_0$, are provided by the linelist, then scaled by their temperature dependence according to semiclassical impact theory (e.g. \citealp{hubenyTheoryStellarAtmospheres2014} ch. 8.3) to the per-particle broadening parameters:
\begin{equation}
    \gamma_\text{stark} = \gamma^\text{stark}_0 \left (\frac{T}{10^5~\text{K}} \right )^\frac{1}{6}
\end{equation}
\begin{equation}
    \gamma_\text{vdW} = \gamma^\text{vdW}_0 \left (\frac{T}{10^5~\text{K}} \right )^\frac{3}{10}
\end{equation}
If provided, ABO parameters, which describe a more nuanced temperature dependence in broadening by neutral hydrogen, will be used to calculate $\gamma_\text{vdW}$ instead.
When a Stark broadening parameter is not provided in the linelist, the approximation from \citet{1971Obs....91..139C} is used.
When a van der Waals broadening parameter is provided, a form of the the Uns\"old approximation \citep{unsoldPhysikSternatmospharenMIT1955, warnerEFFECTSPRESSUREBROADENING1967} is used in which the angular momentum quantum number is neglected and the mean square radius, $\overline{r^2}$, is approximated by
\begin{equation}
    \overline{r^2} = \frac{5}{2} \left(\frac{{n_\mathrm{eff}}^2}{Z}\right)^2 \quad ,
\end{equation}
where $n_\mathrm{eff}$ is the effective principal quantum number and $Z$ is the atomic number.
Pressure broadening is neglected for autoionizing lines with no provided parameters.

The absorption coefficient for each line (except those of hydrogen) is calculated over a dynamically determined wavelength range.  
The maximum detuning (wavelength difference from the line center) is set to the value at which a pure Gaussian or pure Lorentzian profile takes on a value of  $\alpha_\text{crit}$, whichever is larger.
By default, \korg{} truncates profiles at $\alpha_\text{crit} = 10^{-3} \alpha_\text{cntm}$, where $\alpha_\text{cntm}$ is the local continuum absorption coefficient.  
This ratio, $\alpha_\text{crit}/\alpha_\text{cntm}$, can be set by the user.

The broadening of hydrogen lines is treated separately.  
We use the tabulated Stark broadening profiles from \citep{stehleExtensiveTabulationsStark1999}, which are pre-convolved with Doppler profiles.
For H$_\alpha$, H$_\beta$, and H$_\gamma$, where self-broadening is important, we add to $\phi$ a Voigt profile using the p-d approximation for self-broadening from \citet{barklemSelfbroadeningBalmerLine2000}.

\subsection{Radiative transfer} \label{sec:radiative_transfer}
Given the absorption coefficient, $\alpha_\lambda$ at each wavelength and atmospheric layer, the final step is to solve the radiative transfer equation.
The radiative transfer equation is 
\begin{equation}
   \frac{\mu}{\alpha_\lambda} \frac{\dd{I_\lambda}}{\dd{z}}  = S_\lambda - I_\lambda \quad ,
\end{equation}
in a plane-parallel atmosphere, and
\begin{equation}
\frac{\mu}{\alpha_\lambda} \frac{\partial I_\lambda}{\partial r} + 
  \frac{1 - \mu^2}{\alpha_\lambda r} \frac{\partial I_\lambda}{\partial \mu} = 
  S_\lambda - I_\lambda \quad ,
\end{equation}
in a spherical  atmosphere, where $I_\lambda$ is the intensity of radiation at wavelength $\lambda$, $S_\lambda$ the source function, $\alpha_\lambda$ the absorption coefficient, $z$ the negative depth into the atmosphere, $r$, the distance to the center of the star, and $\mu$ the cosine of the angle between the $r$/$z$ axis and the line of sight.
When the thickness of the atmosphere is small relative to the stellar radius, curvature can be neglected and a plane-parallel atmosphere is a good approximation, otherwise sphericity must be taken into account.
By default, \korg{} does it's radiative transfer calculations in the same geometry as the model atmosphere.

What \korg{} actually returns is the disk-averaged intensity, i.e. the astrophysical flux,
\begin{equation} \label{eq:astroflux}
    \mathcal{F}_\lambda = 2\pi \int_0^1 \mu I^\text{top}_\lambda(\mu) \, \dd{\mu} \quad .
\end{equation}
Here, $I^\text{top}_\lambda(\mu)$ stands for $I_\lambda(z=0, \mu)$ in the plane-parallel case and $I_\lambda(r=R, \mu)$ in the spherical case, where $R$ is the radius of the outermost atmospheric layer. The total flux from the star is then $F_\lambda = (R/d)^2 \mathcal{F}_\lambda$, where $d$ is the distance to the star.
Since \korg{} assumes that the stellar atmosphere is in LTE, $S_\lambda$ is the blackbody spectrum.

In the plane-parallel case, 
\begin{equation} \label{eq:plane_parallel_F}
    \mathcal{F}_\lambda = 2 \pi \int_{\tau_\lambda'}^0 S_\lambda(\tau_\lambda) E_2(\tau_\lambda) \dd{\tau_\lambda} \quad ,
\end{equation}
where $E_2$ is the second order exponential integral, $\tau_\lambda$ is the optical depth ($\dd{\tau_\lambda} = \alpha_\lambda \dd{z}$) and $\tau_\lambda'$ is the depth at the bottom of the atmosphere.
The spherical case admits no further analytic simplifications.
When using a spherical model atmosphere, to obtain the astrophysical flux, $\mathcal{F}_\lambda$, \korg{} calculates the intensity, $I^\text{top}_\lambda$ at a discrete grid of $\mu$ values, by integrating along rays from the lowest atmospheric layer to the top atmospheric layer for many of surface $\mu$ values.
This is valid since the source function is isotropic under LTE.
Rays which do not intersect the lowest atmospheric layer are cast from the far side of the star.
The integral over $\mu$ (Equation \ref{eq:astroflux}) is performed using Gauss-Legendre quadrature.

\change{
\korg{} has two radiative transfer calculation schemes: a lower-order default, and the quadratic Bezier scheme from \citet{delacruzrodriguezDELOBezierFormalSolutions2013}.
The agree at the sub-percent level in flux, and complication with the Bezier scheme lead us to make the lower-order scheme the default.
Appendix \ref{ap:transfer} contains numerical details and an extended discussion of the calculation of transfer integrals.
}

\section{Comparison to other codes} \label{sec:verification}
\begin{figure*}
    \centering
    \includegraphics[width=\textwidth]{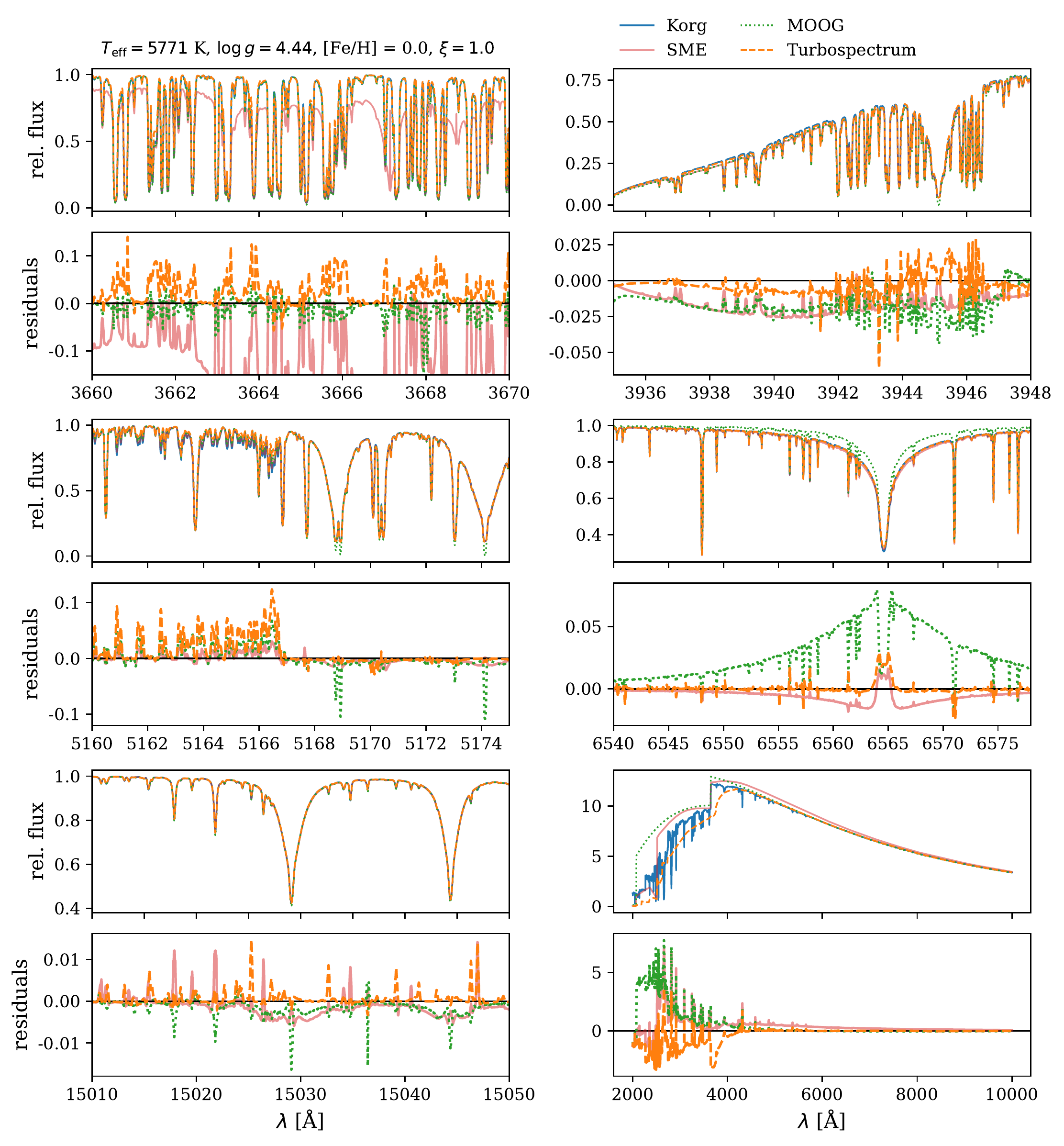}
    \caption{Portions of a synthetic solar spectrum generated with \korg{}, \moog{}, \ts{}, and \sme{}, as well as the continuum generated with an empty linelist from 4000 \AA{} to 10000 \AA. All wavelengths are vacuum.
    We emphasize that the structure in the blue and ultraviolet in the \korg{} continuum is due to resonances in the metal bound-free cross-sections, not lines.
    The residuals (other - \korg{}) are shown underneath the rectified flux for each wavelength region.
    The level of agreement between \korg{} and the other codes is similar to their agreement with each other. See text for a discussion of the discrepancies. 
    }
    \label{fig:sun}
\end{figure*}

\begin{figure*}
    \centering
    \includegraphics[width=\textwidth]{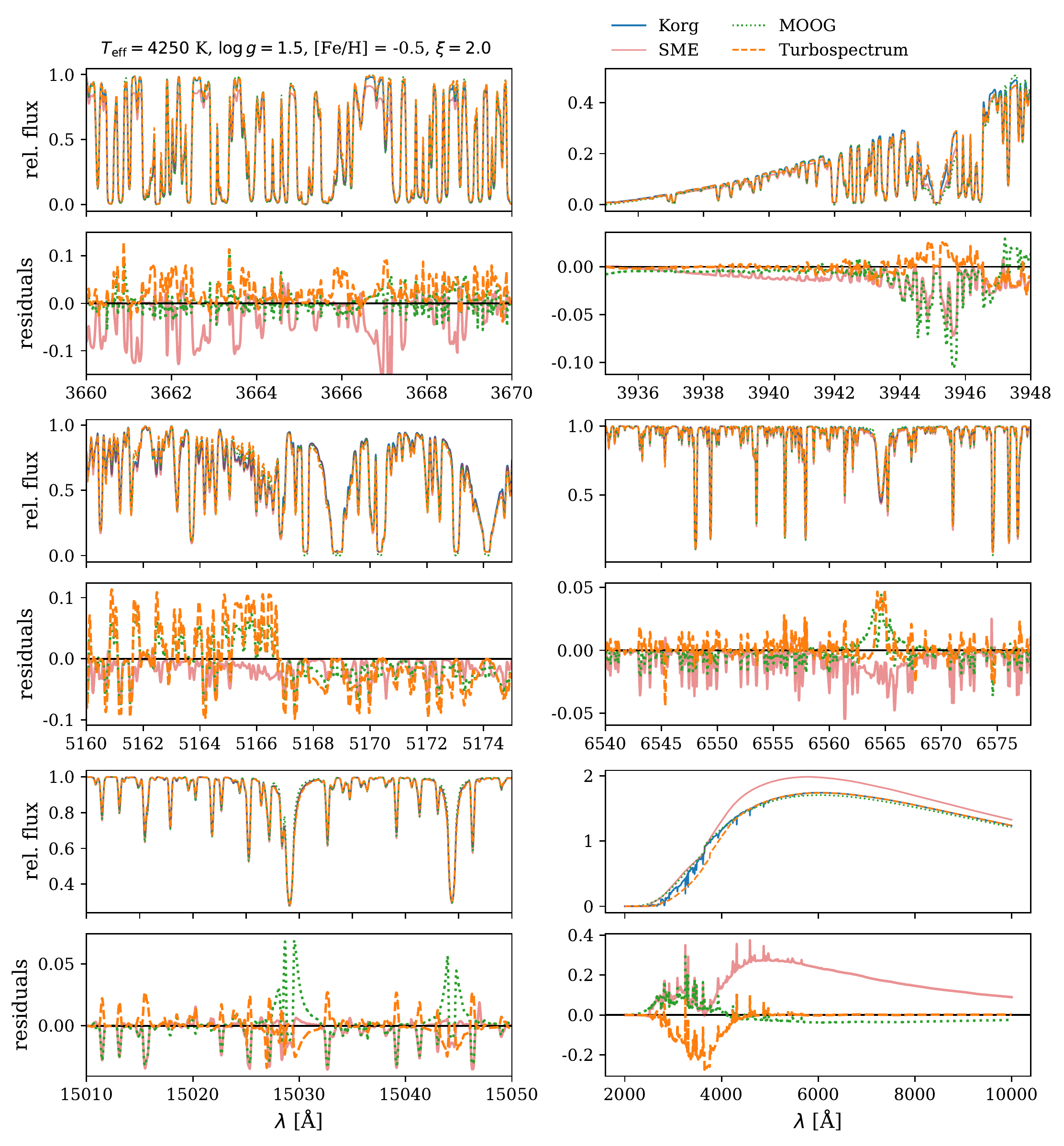}
    \caption{Same as Figure \ref{fig:sun}, but showing the synthetic spectrum of an Arcturus-like star.}
    \label{fig:arcturus}
\end{figure*}

\begin{figure*}
    \centering
    \includegraphics[width=\textwidth]{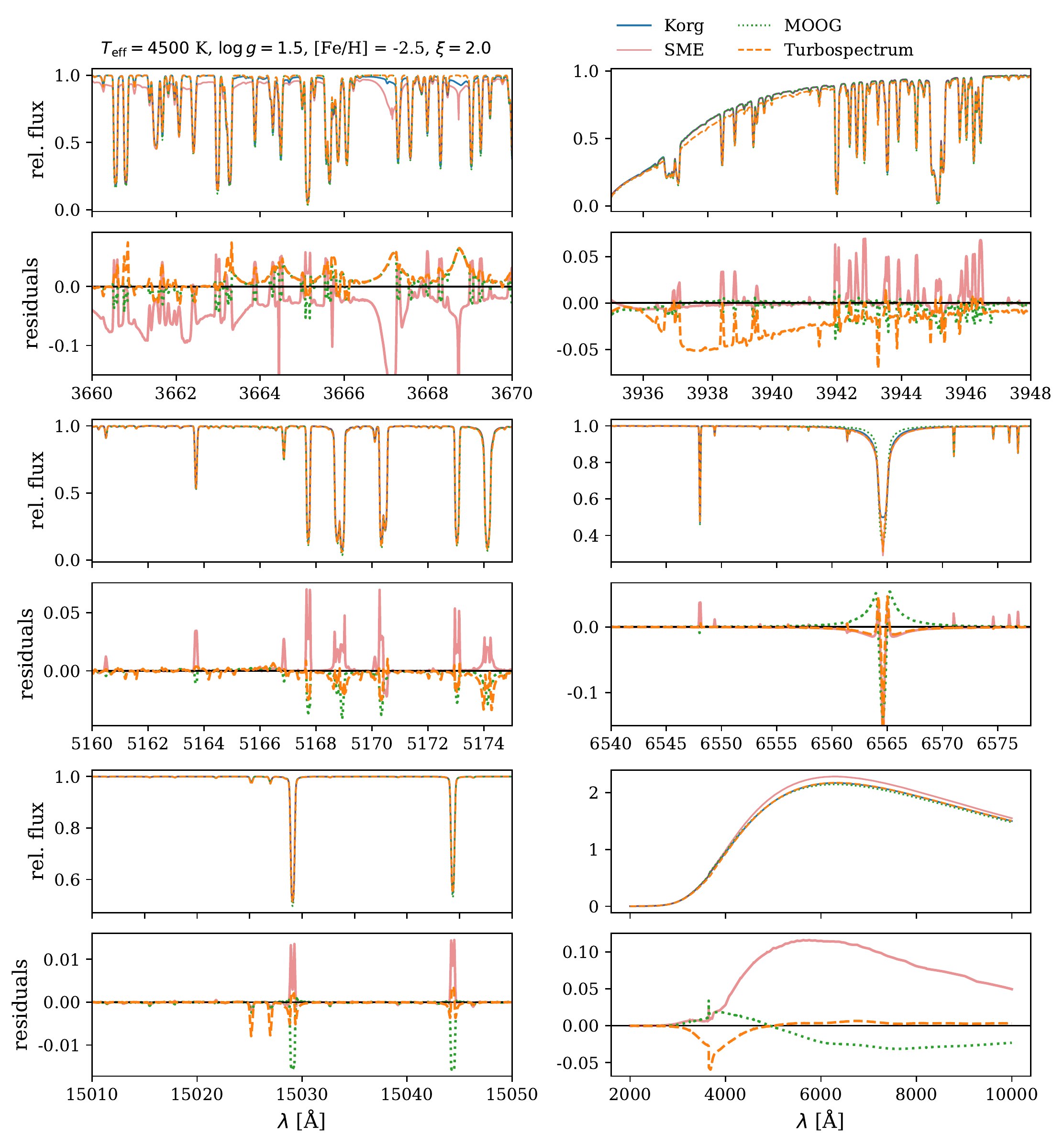}
    \caption{Same as Figure \ref{fig:sun}, but showing the synthetic spectrum of an HD122563-like star.}
    \label{fig:HD122563}
\end{figure*}

\begin{figure*}
    \centering
    \includegraphics[width=\textwidth]{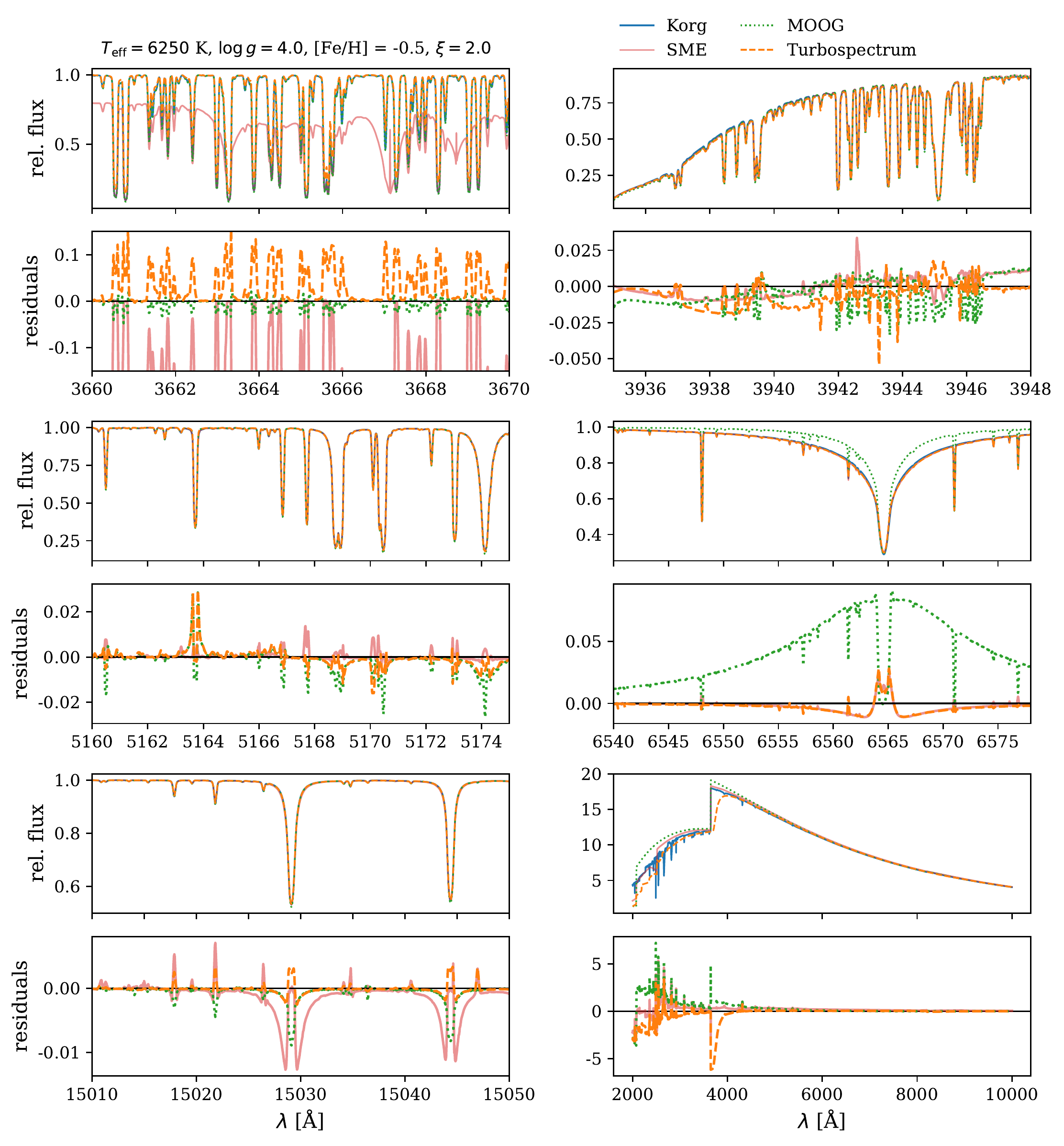}
    \caption{Same as Figure \ref{fig:sun}, but showing the synthetic spectrum of an HD499330-like star.}
    \label{fig:HD49933}
\end{figure*}

To test \korg{}'s correctness, we compare its spectra to those produced by \moog{}, \ts{}, \change{and \sme{}} for four sets of the parameters ($T_\mathrm{eff}$, $\log g$, and metallicity).
\change{The first set of parameters is solar, and we have chosen the others to be similar to three well-studied stars: the red giant branch star Arcturus, HD122563 (an extremely metal-poor red giant), and HD49933 (an F-type main sequence star), but shifted slightly to the nearest existing MARCS model atmosphere to avoid the need to interpolate the atmospheres.}
For all stars, the \change{same} solar abundance pattern was assumed, which is not problematic since we are comparing synthetic spectra to each other and not to observational data.
Figures \ref{fig:sun} -- \ref{fig:HD49933} show these comparisons in six vacuum wavelength regions: 
\begin{itemize}
    \item 3660 \AA{} -- 3670 \AA{}, near the Balmer jump
    \item 3935 \AA{} -- 3948 \AA{}, in the wing of the Ca II Fraunhofer $K$ line
    \item 5160 \AA{} -- 5175 \AA{}, including two lines of the Fraunhofer $b$ Mg triplet 
    \item 6540 \AA{} -- 6580 \AA{}, including $H_\alpha$
    \item 15000 \AA{} -- 15050 \AA{}, in the near-infrared, part of the APOGEE wavelength region
    \item The continuum across 2000\AA{} -- 10000 \AA, computed without any lines
\end{itemize}
For the first four regions, we used an ``extract all'' linelist from VALD, which includes all known lines within the wavelength bounds.
The near-infrared spectrum was synthesized using the APOGEE DR16 linelist \citep{smithAPOGEEDataRelease2021}. 
The VALD linelists are pre-adjusted for isotopic abundances, which eliminates a possible source of disagreement.  
\change{For the APOGEE linelist, \ts{}'s runtime isotopic adjustment may not be consistent with \korg{}'s or with the pre-adjustment performed for \sme{} or \moog{}, but agreement is generally good nevertheless.}
As \moog{} has a limit of 2500 lines which can contribute at a given wavelength, we culled the weakest TiO lines from each list, reducing the total number of transitions passed to \moog{} to be at most 10,000.
This is expected to have negligible impact on the spectrum.
Since \korg{} (see Section \ref{sec:continuum}) and \moog{} (excepting the fork presented in \citealp{sobeckAbundancesNeutroncaptureSpecies2011}) do not support true scattering, we set the \texttt{PURE-LTE} flag to true in \texttt{babsma.par}, setting \ts{} to turn off this functionality.
\change{Scattering can't be turned off in \sme{} which presumably explains some of the disagreement between it and other codes.}

\change{Disagreement between the codes is substantial, with fluxes disagreeing at the ten percent level in many cases, as established by \citet{blanco-cuaresmaModernStellarSpectroscopy2019}.
Agreement is strongest in the infrared, where continuum opacities are simpler.  We discuss some of the discrepancies in more detail here.}


\subsection{Continuum absorption in the violet and ultraviolet}
\change{
Near the Balmer jump and blueward, agreement between the synthetic continua is generally poor.
This is the primary cause for the disagreement between the codes in the first two wavelength regions, which are at the shortest wavelengths.
The defined structure seen the \korg{} continuum is due to resonances in the metal bf cross-sections, not lines.
Unfortunately, it is the case that reliable spectral synthesis in the blue and ultraviolet remains out of reach, as metal opacities are theoretically very challenging to predict at the energy resolution required (not to mention the difficult-to- effects of line blanketing).
}

\subsection{Balmer series} \label{sec:balmer}
\begin{figure*}
    \centering
    \includegraphics[width=\textwidth]{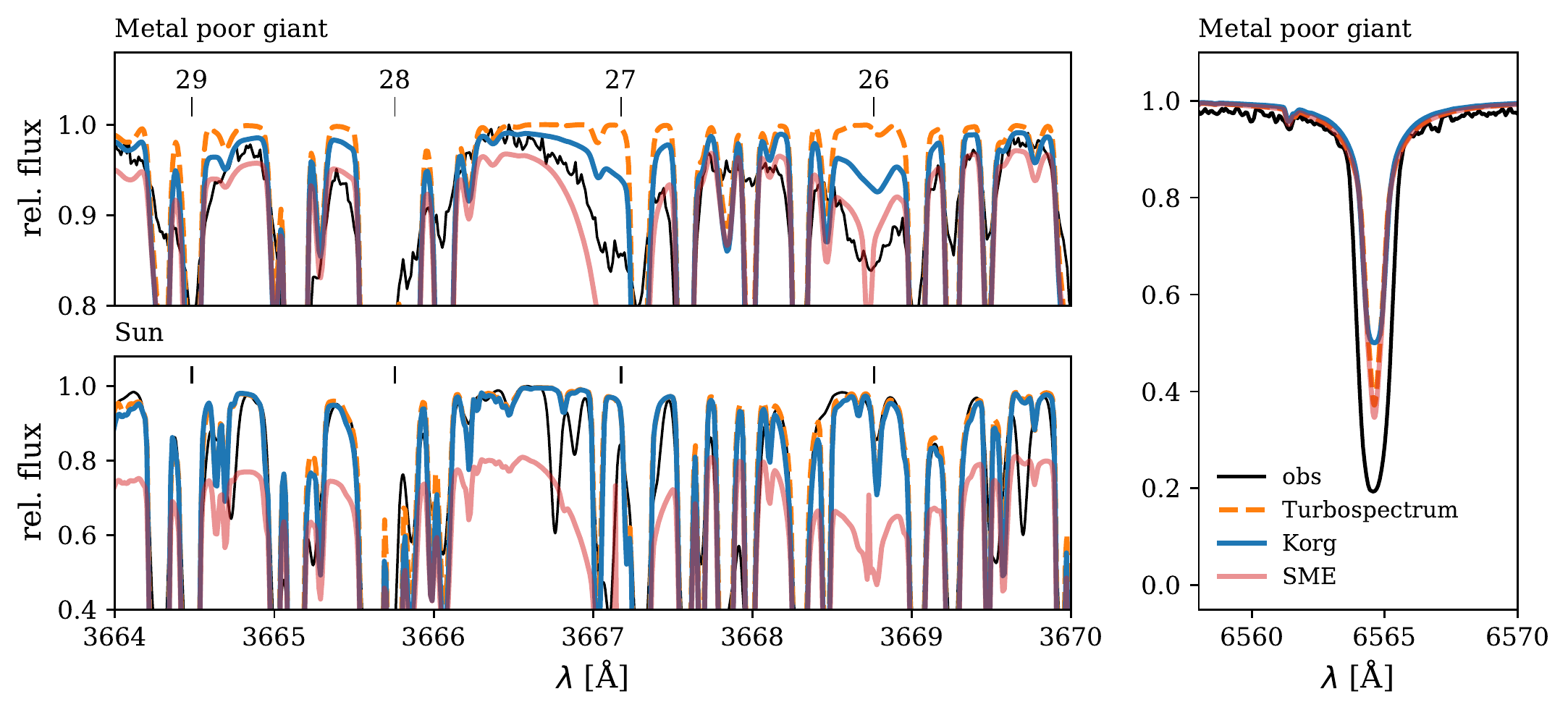}
    \caption{\textbf{top-left:} part of the region with high-order Balmer lines for the metal-poor giant, along with the observed spectrum of HD122563. \textbf{bottom-left:} the same for solar parameters, alongside the observed solar spectrum \textbf{right:} $H_\alpha$ in the metal-poor giant, along with the observed spectrum of HD122563. None of the codes correctly model the hydrogen lines in the metal-poor regime. Though \sme{} is the closest to the observed high-order Balmer lines in the metal-poor regime, it predicts strong high-order Balmer lines in stars where they are not present. All synthetic spectra have been convolved to the observational resolution.}
    \label{fig:balmer}
\end{figure*}
There are eight Balmer series lines in the 3660 \AA{} -- 3670 \AA{} (vacuum) wavelength region (upper levels $n=26$ to $n=33$), five of which are included in \korg{} (those up to $n=30$).
Typically, these have a minimal impact on the observed spectrum, but in the metal-poor giant, \korg{} predicts that these lines are visible as broad shallow features, in contrast to \ts{} (hydrogen lines are omitted from the \moog{} linelist \change{in this region}).
They are located at 3668.76 \AA, 3667.17 \AA, 3665.75 \AA, 3664.48 \AA, and 3663.33 \AA{} (vacuum), and are more clearly visible in the residuals panel. 
\change{
This is because \ts{} uses the occupation probability formalism discussed in Section \ref{sec:moleq}, which eliminates the relevant hydrogen orbitals throughout most of the stellar atmospheres.
}
In a high-resolution ultraviolet spectrum of HD122563 from the Ultraviolet and Visual Echelle Spectrograph\footnote{ \href{http://archive.eso.org/dataset/ADP.2021-08-26T17:20:56.312}{this spectum} (archive ID: ADP.2021-08-26T17:20:56.312).} \citep{dekkerDesignConstructionPerformance2000}, a star with similar parameters, the lines are present.
\change{
For this reason, we are not confident that the occupation probability formalism offers a significant advantage over unmodified partition functions, and we do not use it in \korg{}.
The top-left panel in Figure \ref{fig:balmer} shows part of this wavelength region in more detail, along with the observational data.
\sme{} predicts the strongest absorption by these lines in the metal-poor  giant, but it also predicts strong absorption for the other stars, where it is not present in observations.
The bottom-left panel of Figure \ref{fig:balmer} demonstrates this for the Sun, using the \citet{wallaceOPTICALNEARINFRARED295892502011} solar atlas.
}

\change{There is relatively good agreement between \korg{}, \ts{}, and \sme{} in the $H_\alpha$ wings (\moog{} lacks special treatment of hydrogen lines, and generally disagrees).
\ts{} and \sme{} both use forms of \texttt{HLINOP} \citep{barklemHydrogenBalmerLines2003}, so strong agreement is to be expected (though they presumably have differing treatments of the equation of state).
\newchange{\korg{} uses roughly the same treatment as \texttt{HLINOP}, with Stark broadening from \citet{stehleExtensiveTabulationsStark1999}  self-broadening via the Voigt approximation from \citet{barklemHydrogenBalmerLines2003}, but Figure \ref{fig:balmer} indicates that there are differences in implementation.}
For the metal-poor giant hot dwarf (Figures \ref{fig:HD122563} and \ref{fig:HD49933}), \ts{} and \sme{} agree with each other almost exactly, and disagree with \korg{} at the $\sim1\%$ level in the $H_\alpha$ wings.
In constrast, \korg{} and \ts{} agree more closely with each other than they do with \sme{} for the Sun.
In the $H_\alpha$ core, disagreement is similarly minor, except for in the metal-poor giant.
The right panel in Figure \ref{fig:balmer} shows this in detail, alongside the observed spectrum of HD122563, a star with similar parameters, from GALAH \citep{buderGALAHSurveyThird2021}.
Given that accurate modelling of the $H_\alpha$ core requires techniques beyond 1D LTE (e.g. \citealp{barklemNonLTEBalmerLine2007, amarsiEffectiveTemperatureDeterminations2018}), and that the observational data doesn't match the prediction of any of the codes, we consider this disagreement permissible.
}

\subsection{C$_2$ band} \label{sec:C2}
\begin{figure*}
    \centering
    \includegraphics[width=\textwidth]{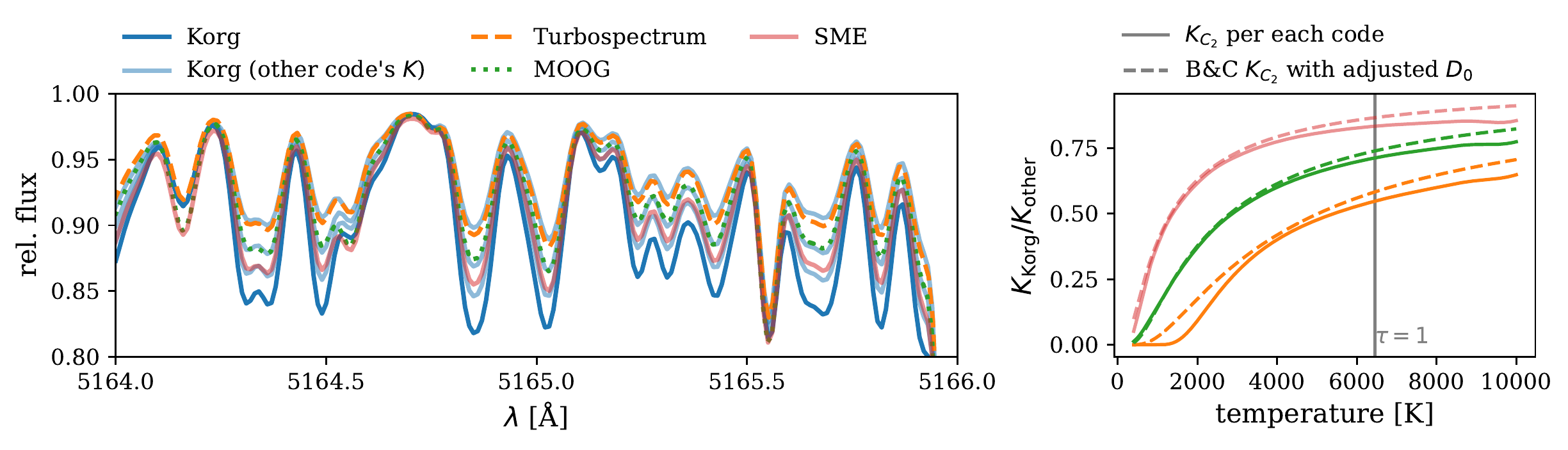}
    \caption{\textbf{left:} Some lines from a C$_2$ band in the sun synthesized by the four codes, and synthesized by \korg{} using the prescriptions for the chemical equilibrium constant, $K_{\mathrm{C}_2}$ from the other codes.  Differences in these prescriptions, primarily arising from differences in the dissociation energy, $D_0$, drive most of the disagreement. \textbf{right:} the ratio of $\kcc$ values per \citet{barklemPartitionFunctionsEquilibrium2016} (\korg{}'s adopted values) and per the treatment used by another code.
    The dashed lines show the values obtained using the \citet{barklemPartitionFunctionsEquilibrium2016} data adjusted for the $D_0$ used by another code.  Uncertainty in $D_0$ drives most of the disagreement in the amount of $C_2$ present at the photosphere and most of the disagreement in the synthesized spectra at these wavelengths.
    }
    \label{fig:kc2}
\end{figure*}
\change{
While tracking down the causes of disagreement between codes for all wavelengths and stars isn't feasible, it is instructive to focus on one as a ``case study''.
In the sun, \korg{} predicts deeper lines from roughly 5160 \AA{} -- 5165 \AA{} than the other codes.
These are due to absorption by C$_2$, and the disagreement arises from the varying molecular equilibrium constants, 
\begin{equation}
    \kcc = \frac{P_\mathrm{C}^2}{P_{\mathrm{C}_2}}  \quad ,
\end{equation}
adopted for the species by the codes.
Most of the difference in $\kcc$ values comes from different values for the dissociation energy, $D_0$, of C$_2$.
Molecular dissociation energies are one of the most dominant sources of systematic uncertainty in spectral synthesis (see discussion in \citealp{barklemPartitionFunctionsEquilibrium2016}).
To summarize the treatments of $\kcc$ in each code:
\begin{itemize}
\item \korg{}  uses the data from \citet{barklemPartitionFunctionsEquilibrium2016}: $D_0 = 6.371$ eV.
\item Based on logging information, \ts{} appears to use the polynomial expansions from \citet{tsujiMolecularAbundancesStellar1973}: estimated $D_0 = 6.07$ eV.
\item The \moog{} molecular equilibrium constants comes from Kurucz\footnote{\url{http://kurucz.harvard.edu/molecules.html}}. We were able to extract the polynomial approximation for the C$_2$ molecular equilibrium constant from the source code: $D_0 = 6.21$ eV.
\item For C$_2$, the \sme{} equilibrium constant comes from \citet{sauvalSetPartitionFunctions1984}: $D_0 = 6.297$ eV.
\end{itemize}
}

\change{
Figure \ref{fig:kc2} shows the effects of these choices about $K_{\mathrm{C}_2}$.
They account for nearly all of the disagreement in the synthesized spectra.
While the numerical scheme used to represent the temperature-dependence of $\kcc$ plays some role, the value of $D_0$ adopted is far more important.
The differing $D_0$ values result in number densities of C$_2$ differing by up to a factor of 2 at the photosphere, and by orders of magnitude at the top of the atmosphere.
}

\section{Benchmarks} \label{sec:benchmarks}
\begin{figure}
    \centering
    \includegraphics[width=0.45\textwidth]{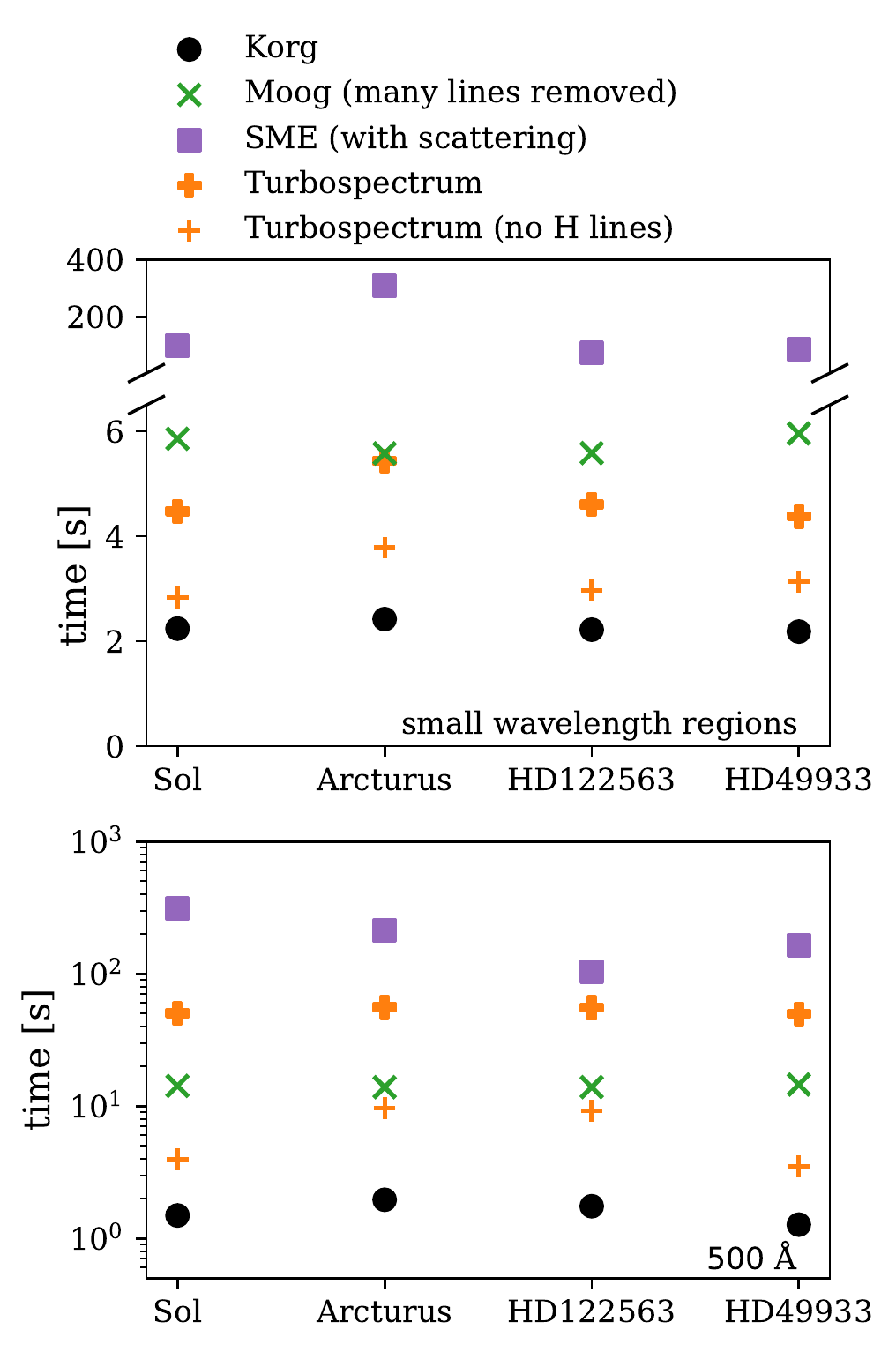}
    \caption{The time taken by each code for the syntheses in Section \ref{sec:verification}.
    \textbf{top: } average compute time for the optical wavelength regions, which used very dense linelists.
    \textbf{bottom: } compute time to synthesize spectra from 15,000 -- 15,500\,\AA{} with the APOGEE DR16 linelist, a more ``realistic'' use case.
    Note that the wavelength regions synthesized were larger than those plotted in Figures \ref{fig:sun} -- \ref{fig:HD49933}. 
    \change{\sme{} includes a true scattering treatment which can't be turned off, increasing its execution times.}
    }
    \label{fig:benchmark}
\end{figure}

The first panel in Figure \ref{fig:benchmark} shows the time taken by each code to produce the spectra in the first four wavelength regions in Section \ref{sec:verification}. 
These are relatively small regions with linelists including all transitions in VALD, including many which have essentially no effect on the spectrum.
The second panel shows the time to compute a spectrum from 15000 \AA{} -- 15500 \AA{} using the APOGEE DR16 linelist.
This provides a more realistic example of synthesizing a spectrum as part of the analysis for a large survey.
All test are single-core, and run on the same machine with an AMD Epyc 7702P.

\korg{} only needs to load the linelist and model atmosphere into memory once, so repeat syntheses which use the same inputs (as would be the case when varying individual abundances) avoid that step.
For these comparisons, loading the linelist ($10^5$ -- $10^6$ transitions) takes 1 -- 4 seconds, and loading the model atmosphere takes roughly 0.05 ms.
As nearly all cases requiring performance involve repeated synthesizing with the same (or nearly the same) linelist, we did not optimize input parsing for speed and do not include this time in \korg{} benchmarks.
We note that for the ``small wavelength regions'', the \moog{} times are technically lower limits, since the linelists provided to \moog{} were reduced in size by an order of magnitude or more (though the synthesized spectra are essentially unchanged).
Figure \ref{fig:benchmark} has separate markers for \ts{} with and without hydrogen lines, as we found them to slow down synthesis by a very large factor.
In wavelength ranges where hydrogen lines are not present or unimportant, they can be omitted, and \ts{} executes much more quickly.
\change{The comparison to \sme{} is included for completeness, but is not entirely fair since \sme{} includes the effects of scattering, which is numerically expensive.
Scattering is turned off for \ts{}, to make the comparison as fair as possible.}

\begin{figure}
    \centering
    \includegraphics{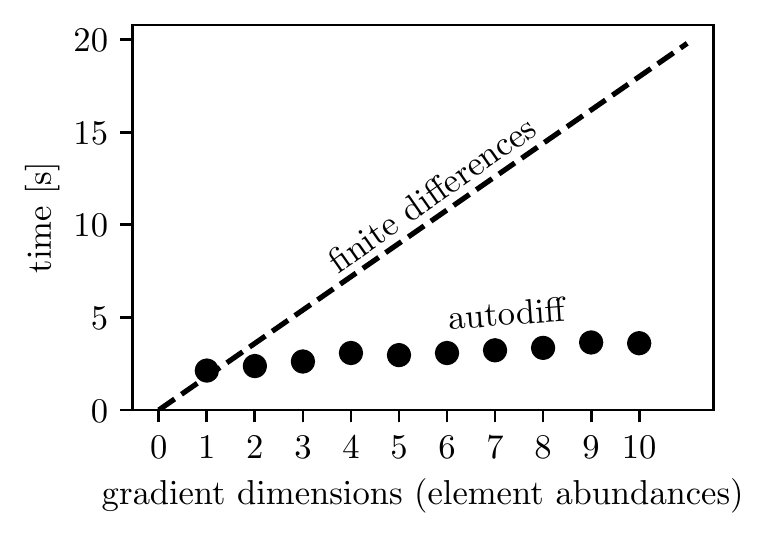}
    \caption{\textbf{top:} Time required to compute simultaneous derivative of spectrum with regard to $N$ different element abundances, for varying $N$.}
  \label{fig:diff_benchmark}
\end{figure}

In the top panel of Figure \ref{fig:diff_benchmark} we plot the time required by \korg{} to compute the gradient of the solar spectrum, $\frac{\partial \mathcal{F}}{\partial A(X)}$, with regard to a varying numbers of element abundances, $N$.
(See Figure \ref{fig:gradients} for a subset of the derivative spectra.)
For this demonstration we used the 15,000 -- 15,500\,\AA{} range and the APOGEE DR16 linelist.
A dashed horizontal line marks the time required to synthesize the spectrum, just under one second.  
Calculating the $N$-dimensional gradient spectrum takes roughly $2 + 0.15N$ seconds, meaning that the marginal time required for each derivative is an order of magnitude smaller than the time required to obtain it via finite differences. 
As \textsc{Julia}'s automatic differentiation ecosystem improves, \korg{} may benefit from further speedups to the calculation of gradient spectra.
Calculating derivative spectra with respect to atmospheric parameters, e.g. $T_\mathrm{eff}$, $\log g$, or $v_\mathrm{mic}$, requires code to interpolate between model atmospheres, which we plan to add to \korg{}.

\section{Conclusions and future development}
We have presented a new code, \korg{}, for 1D LTE spectral synthesis, i.e. computing stellar spectra given a model atmosphere, linelist, and element abundances.
\change{\korg{} is both fast faster than all other codes, and autodifferentiable, which yields further speedups when synthesis is embedded in an optimization loop.}
The code is publicly available at \url{https://github.com/ajwheeler/Korg.jl} and installable via the \textsc{Julia} package manager.
Detailed documentation, usage examples, and installation instructions are available at \url{https://ajwheeler.github.io/Korg.jl/stable}.

\change{
In comparing \korg{} to other codes, we have highlighted that the level of disagreement between them (and thus systematic error) is substantial.
Much of this can be attributed to uncertain physical parameters (e.g. the C$_2$ dissociation energy, discussed in Section \ref{sec:C2}) and models of continuum absorption processes, but varying numerical methods also play a role (e.g. the discussion in appendix \ref{ap:transfer}).
Unfortunately, similar problems extend to model atmospheres, which require the same physics used for synthesis, and linelists, which include many poorly-constrained atomic parameters.
}

\change{
We have shown in this work that \korg{} produces results for FGK stars that are similar to other codes. We recommend its use for these spectral classes.
We plan to soon address the factors limiting \korg{}'s applicability outside this regime, most importantly its lack of polyatomic molecules (present in cool stars) and true scattering (important in hot stars).
We also plan to add support for NLTE departure coefficients.
}
There are also limitations affecting ease-of-use that we plan to address.
As mentioned in Section \ref{sec:benchmarks}, calculating derivative spectra with regard to atmospheric parameters like $\log g$ or $T_\mathrm{eff}$, requires code to interpolate model atmospheres which is written in \textsc{Julia}.
Relatedly, while inferring stellar abundances and parameters with \korg{} is not overly burdensome, the user must still apply the line spread function to synthesized spectra and calculated goodness-of-fit themselves, processes we would like to automate.

\change{
There are some limitations which apply to all spectroscopic synthesis codes.
In the blue and ultraviolet, the poorly-constrained continuum results in very uncertain predictions, manifesting in the poor agreement found between codes.
Many of the chemical equilibrium constants necessary for spectroscopic synthesis are also poorly-constrained, as discussed \citet{barklemPartitionFunctionsEquilibrium2016}.
There are doubtless many features (like the one discussed in Section \ref{sec:C2}) where chemical equilibrium constants are the primary driver of uncertainty.
Uncertain atomic parameters in linelists pose a similar problem in all regimes.
}


Our goal is that \korg{} will be useful for both inference of stellar parameters and abundances for large survey data, e.g. \emph{Gaia} \citep{gaiacollaborationGaiaMission2016}, SDSS-V \citep{kollmeierSDSSVPioneeringPanoptic2017}, MOONS \citep{kollmeierSDSSVPioneeringPanoptic2017}, WEAVE \cite{daltonWEAVENextGeneration2012}, 4-MOST \citep{dejong4MOSTProjectOverview2019}, LAMOST \citep{dengLAMOSTExperimentGalactic2012, zhaoLAMOSTSpectralSurvey2012}, GALAH \citep{desilvaGALAHSurveyScientific2015}, and for boutique analyses of individual spectra.
We aim to make \korg{} fast and flexible enough to enable better survey pipelines and novel analyses, such as the propagation of error from synthesis inputs to synthesized spectra, or the joint inference of line parameters with observational data.
In addition, we hope that by making \korg{} as easy to use as possible, more researchers will find it worthwhile to synthesize spectra when the need arises.

\section*{Acknowledgments}
AJW would like to thank Chris Sneden for his advice, Samuel Potter, Paul Barklem, Karen Lind, and Thomas Nordlander for answers to naive questions, and Rob Rutten, David F. Gray, Ivan Hubeny, Robert Kurucz, and Dimitri Mihalas for writing excellent reference material. He would also like to thank Charlie Conroy for his interest and encouragement, and David Hogg for putting him in touch with Mike O'Neil and Samuel Potter. The authors would like to thank the anonymous reviewer for their thoughtful comments and expertise.

AJW is supported by the National Science Foundation Graduate Research Fellowship under Grant No. 1644869. MKN is in part supported by a Sloan Research Fellowship. A.~R.~C. is supported in part by the Australian Research Council through a Discovery Early Career Researcher Award (DE190100656) \change{and through a Monash University Network of Excellence grant (NOE170024)}. Parts of this research were supported by the Australian Research Council Centre of Excellence for All Sky Astrophysics in 3 Dimensions (ASTRO 3D), through project number CE170100013.

This research has made use of the services of the ESO Science Archive Facility.
Based on observations collected at the European Southern Observatory under ESO programme 266.D-5655.
This work has made use of the VALD database, operated at Uppsala University, the Institute of Astronomy RAS in Moscow, and the University of Vienna.
The authors acknowledge support and resources from the Center for High-Performance Computing at the University of Utah.


\bibliography{Korg}

\appendix

\section{Numerical details of radiative transfer} \label{ap:transfer}
\subsection{\korg{}'s default implementation}
To solve the equation of radiative transfer along a ray, \korg{} approximates the source function, $S$ ($\lambda$ subscripts dropped for brevity), with linear interpolation over optical depth, $\tau$.  Between each adjacent pair of atmospheric layers, we have $S(\tau) \approx m\tau + b$, so the (indefinite) transfer integral has solution
\begin{equation}
    \int (m \tau - b) \exp(-\tau) \mathrm{d} \tau = -\exp(-\tau) \left(b + m(\tau + 1)\right) \quad .
\end{equation}
When evaluating equation \ref{eq:plane_parallel_F}, \korg{} uses the same approximation, 
\begin{equation}
    \int (m \tau - b) E_2(\tau) \mathrm{d} \tau = \frac{1}{6} \left[\tau E_2(\tau) (3b + 2m\tau) - \exp(\tau) (3b + 2m(\tau + 1))\right ] \quad .
\end{equation}
Calculating the optical depth, $\tau$, requires numerically integrating 
\begin{equation}\label{eq:naive_tau}
    \tau_\lambda(s) = \int_0^s \alpha_\lambda \frac{\alpha'_5}{\alpha_5}\dd{s}\quad ,
\end{equation}
where $\alpha_5$ is the absorption coefficient at the reference wavelength, 5000 \AA, calculated by \korg{}, and $\alpha'_5$ is the absorption coefficient at the reference wavelength supplied by the model atmosphere. 
Their ratio is the correction factor to $\alpha_\lambda$ which enforces greater consistency between the model atmosphere and spectral synthesis.
\korg{} computes $\tau_\lambda$ by evaluating the equivalent integral
\begin{equation}\label{eq:fancy_tau}
    \tau_\lambda(s) = \int_0^{\ln \tau'_5(s)}  \tau'_5 \frac{\alpha_\lambda}{\alpha_5} \dd{(\ln \tau'_5}) \quad ,
\end{equation}
where $\tau'_5$ is the optical depth at the reference wavelength per the model atmosphere.
This integral is numerically preferable, since atmospheric layers are spaced uniformly in $\ln \tau'_5$, and the integrand of \ref{eq:fancy_tau} is more nearly linear than that of \ref{eq:naive_tau}.

\subsection{Comparison to quadratic Bézier scheme}
\begin{figure*} \label{fig:formalsols}
    \centering
    \includegraphics[width=\textwidth]{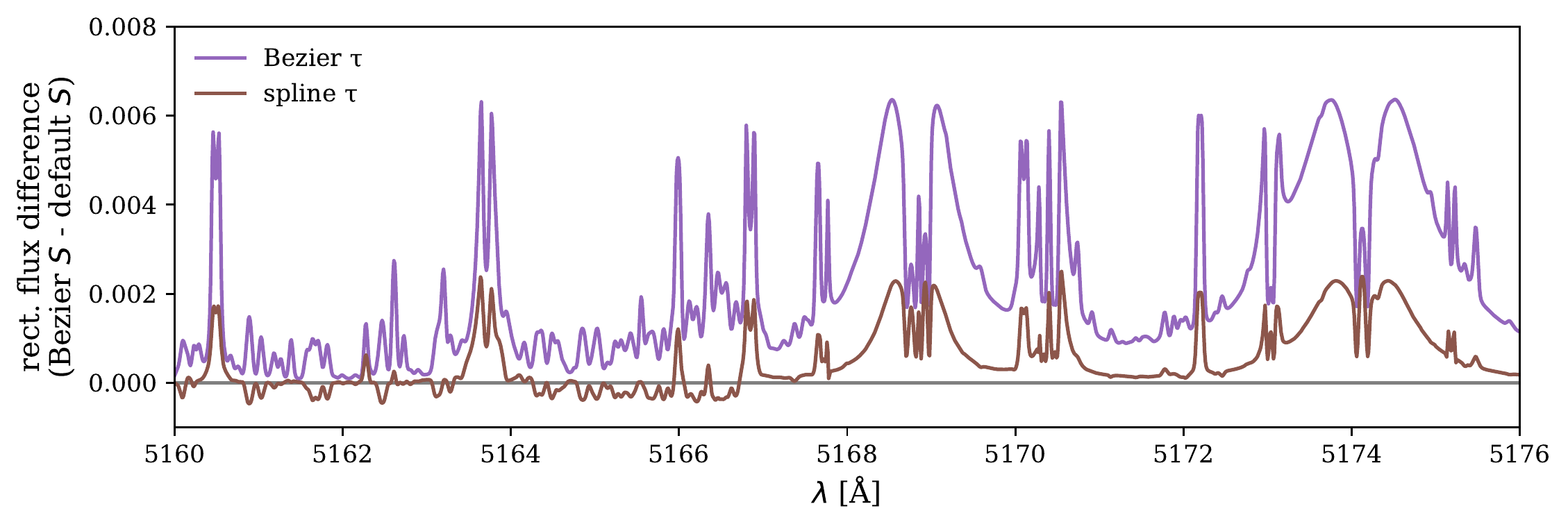}
    \caption{The difference in the rectified spectrum obtained for the Sun using the quadratic Bézier scheme and \korg{}'s default scheme.
    The purple and brown lines show the result obtained using modified Bézier interpolation, and cubic spline interpolation of $\alpha_\lambda$, respectively.
    (This is separate from the interpolation of the source function, which uses Bézier curves in both cases.)
    The difference between all three methods is small compared to the difference in spectra obtained from different codes, but nevertheless larger than is ideal.}
    \label{fig:my_label}
\end{figure*}

\change{
\korg{} has a secondary radiative transfer scheme based on \citet{delacruzrodriguezDELOBezierFormalSolutions2013}, which approximates the source function with monotonic quadratic Bézier interpolation.
They suggest that $\tau_\lambda$ might be computed using the same scheme, but this causes numerical problems when the absorption coefficient is non-monotonic in depth.
We modified the interpolant by requiring the Bézier control point (equation 10 in \citealp{delacruzrodriguezDELOBezierFormalSolutions2013}) to be within a reasonable range, which eliminates the issue, but is physically unmotivated.
This is the implementation available in \korg{}.
As an alternative, we tried computing $\tau_\lambda$ by interpolating $\alpha_\lambda$ with a cubic spline.
Unfortunately, the relatively sparse sampling of model atmospheres means that this choice has an impact on the synthesized spectrum similar to the difference between the default method and the \citet{delacruzrodriguezDELOBezierFormalSolutions2013} method.
Figure \ref{fig:formalsols} shows the difference in a portion of the solar spectrum resulting from using Bézier interpolation of the source function compared to the default implementation.
The difference in synthesized spectra is at the sub-percent level, significantly smaller than the level of disagreement between codes.
While we plan to adopt a higher-order scheme as the default in the future, we provide this scheme as a secondary option until we better understand the impact of the choices involved.
}

\end{document}